\documentstyle[aps,multicol,epsfig]{revtex}
\begin{document}
%\draft

\title{ Magnetic properties of nanoparticles in the Bethe-Peierls approximation}
\author{Luis G. C. Rego and Wagner Figueiredo}
\address{Depto. de F\'{\i}sica, Universidade Federal de Santa Catarina, Florian\'opolis, SC, 88040-900, Brazil }
%\date{\today}
\maketitle
\begin{abstract}
In this work we present a new method to calculate the classical magnetic properties of single-domain nanoparticles. Based 
on the Bethe-Peierls (pair) approximation, we developed a simple system of equations for the classical magnetization of spins 
at any position within the nanoparticle. Nearest neighbor pair correlations are treated exactly for Ising spins, and the 
method can be generalized for various lattice symmetries. 
The master equation is solved for the Glauber dynamics (single-spin-flip) in order to obtain the time evolution of the magnetization.
The capabilities of the model are demonstrated through nontrivial calculations of hysteresis loops as well as field cooling (FC) and 
zero field cooling (ZFC) magnetization curves of heterogeneous non-interacting nanoparticles.
The present method automatically incorporates the temperature and could be adapted to describe an ensemble of interacting nanoparticles.
\end{abstract}

\begin{multicols}{2}

\section{Introduction}

The properties of magnetic nanoparticles and fine particle assemblies have been attracting the 
interest of physicists for many decades.  Below a critical size such nanoscopic magnetic systems develop new 
properties that cannot be found in macroscopic systems, where the material is divided into 
domains in order to decrease the magnetostatic energy. Stoner and Wohlfarth \cite{Stoner} laid the theoretical 
basis, which are widely used nowadays, to describe such low dimensional systems and predicted new effects like the superparamagnetism. 
According to the Stoner-Wohlfarth's model, the strong exchange interaction aligns all the particle spins so that the magnetization inside the 
nanoparticle is assumed to be a uniform field, 
which rotates coherently due to the presence of an external magnetic field. 
Despite being a suitable assumption for homogeneous systems, it is not appropriate to describe 
heterogeneous particles, for instance, formed by regions of localized moments having distinct magnetic character (ferromagnetic,
ferrimagnetic, and antiferromagnetic). 
For such heterogeneous systems the exchange energy plays a central role in determining the magnetization of the 
nanoparticle \cite{Salah,Beam,Kodama}. 
Unfortunately, the non-trivial interaction among the spins is responsible for a highly correlated behavior that is very difficult to describe 
in terms of analytical expressions or exact numerical solutions, for a general case. 
In addition, the boundary conditions imposed by the finite size 
of the nanoparticles and surface disorder complicate otherwise simpler solutions for infinite systems.

In most experiments the magnetization due to an ensemble of nanoparticles is the measured parameter. In this case the long range dipolar interaction
is dominant; it produces a collective energy barrier for the magnetization reversal of a nanoparticle and, consequently, sets a new time 
scale for irreversible processes \cite{Prozorov}. 
The magnetization of independent nanoparticles being experimentally accessible  in the limit of very dilute particle ensembles. However, a series 
of recent experiments, using a micro-SQUID technique, have demonstrated the possibility of measuring both the magnetization reversal mechanism and dynamics of 
individual homogeneous nanoparticles (10-20 nm in diameter) \cite{Wernsdorfer1} and cobalt nanoclusters (3-5 nm in diameter) \cite{Wernsdorfer2}.
The formalism presented in this paper should provide a theoretical foundation for experiments made on single (or independent) heterogeneous nanoparticles.

Magnetic systems are well known for developing long range interactions, 
specially in the continuous phase transitions, where the correlation length is infinite. Because of that 
the calculation of critical exponents require sophisticated theoretical formalisms such as renormalization group 
techniques or high temperature series expansions, or numerical methods such as the Monte Carlo. 
On the other hand, away from the critical point, where the correlation length decays exponentially,
a mean field theory may provide a good description of the behavior of the system. Nevertheless, the plain 
mean field approach corresponds to one in which there is no correlation between any pair of spins  
and a given spin interacts with the average magnetization of its neighbors or, equivalently, with an effective
mean field produced by all the spins in the system. This simplified approach might fail however for nanoparticles 
having a core with magnetic properties that differ from those of the surface, as in the instance of a ferromagnetic 
particle coated by a ferrimagnetic or antiferromagnetic layer \cite{Beam}, caused by oxidation
processes, or the case of a disordered surface as well \cite{Kodama}. 
For these systems the correlations between neighbors are important to establish boundaries between regions of different magnetic 
properties.

In this work we propose a new method to describe the classical magnetic properties of non-interacting nanoparticles. The method takes a further step 
from the standard mean field approach and treats the nearest neighbor pair correlations between spins within the particle explicitly, 
therefore  being capable of describing 
heterogeneous systems.  In spite of being extensively used to calculate the magnetic properties of infinite systems, the Bethe-Peierls 
(also called pair) approximation has not yet been used to calculate the magnetic properties of nanoparticles.
A somewhat similar approach was used to describe the magnetization of Fe quasi-1D clusters
\cite{Kirschner}, in view of the fact that the results obtained by the Bethe-Peierls approximation agree with the Ising solution for 1D 
systems \cite{Plischke}.  
In the following the details of the model and the formalism are presented. Then, in the remainder of the paper, it is shown that the
model accounts for many of the properties of magnetic nanoparticle systems.

\section{Theory} 
\label{theory}

At first, consider a two-dimensional (2D) array of Ising spins in an hexagonal lattice, as 
depicted in Figure 1 for a structure made up of 3 shells. 
A similar model system could be used to describe the properties of single-crystalline disks with hexagonal contour, as the 10-20 nm BaFeCoTiO 
nanoparticles whose magnetization reversal has been measured \cite{Wernsdorfer1}.
We note that the formalism applies equally well for any 2D lattice and can be extended to
tridimensional (3D) lattices as well.
Taking advantage of the sixfold rotational symmetry of the lattice, we need to consider only a small subset of 
spins (marked $\sigma_1$ through $\sigma_6$ by
filled dots in Fig. 1) to provide the  magnetization of the whole particle. Each of the spin sites, which stand for the atomic moments in the nanoparticle, 
are described by Ising spin variables that take on values $\sigma_i = \pm 1$.  
The nanoparticle is assumed to be very small, so that it constitutes a monodomain. In addition,  according to experimental results \cite{Wernsdorfer1}, an uniaxial 
anisotropy is implicit in the model, with the easy axis coinciding with the Ising spin direction. Therefore the energy of the system is written as 
\begin{eqnarray}
\cal{H}  &=&   - \sum_{<i,j>} J_{ij} \sigma_i \sigma_j - H \sum_i \sigma_i \ ,
\label{hamiltonian}
\end{eqnarray}
where only the exchange coupling between nearest neighbors is considered, with $J_{ij}$ as the exchange coupling constant, 
and $H$ is the applied magnetic field.

Having defined the relevant spins that describe the system we start by writing the dynamical
equations for the average spin  magnetizations
\begin{eqnarray}
m_i(t) = <\sigma_i> = \sum_{\sigma} \sigma_i P(\sigma,t)
\end{eqnarray}
and the pair correlation functions
\begin{eqnarray}
r_{ij}(t) = <\sigma_i \sigma_j> = \sum_{\sigma} \sigma_i \sigma_j P(\sigma,t),
\end{eqnarray}
where the index $i$, as well as $j$,  designates a spin site and the sum over $\sigma$ accounts for all the spin 
configurations. Since we restrict our model to take into account only pair correlations, the probability of a given spin configuration 
is written as \cite{Glauber}
\begin{eqnarray}
P(\sigma,t) = \frac{1}{2^N} \left\{ 1 + \sum_i \sigma_i m_i(t) + \sum_{i<j} \sigma_i \sigma_j r_{ij}(t) \right\}  \ ,
\label{prob}
\end{eqnarray}
where $N$ is the total number of spins.

The time evolution of the system is obtained by means of the master equation for the probabilities
\begin{eqnarray}
\frac{dP(\sigma,t)}{dt} = \sum_{\sigma'} \left\{ P(\sigma',t) w(\sigma',\sigma) - P(\sigma,t) w(\sigma,\sigma') \right\} ,
\label{master}
\end{eqnarray}
with $w(\sigma',\sigma)$ as the transition probability per unit time from the spin configuration $\sigma'$ to configuration $\sigma$. 
Here we assume that
the dynamics of the system is governed by single-spin-flip transitions, which can be mathematically described by the Glauber formalism
\cite{Glauber}
\begin{eqnarray}
w_i(\sigma_i) = \frac{1}{2\tau} \left\{ 1 - \sigma_i \gamma \right\} \ ,
\label{w}
\end{eqnarray}
with 
\begin{eqnarray}
\gamma(J,\beta,H) &=& \tanh{[(\sum_{j} J_{ij} \sigma_j + H) / k_B T]} \\ 
&\equiv&  \tanh{(\Delta_{i})} , 
\label{gamma}
\end{eqnarray}
where $k_B$ is the Boltzmann constant, $T$ is the absolute temperature  and $\tau$ is the relaxation time for a single spin.
This dynamical rule is chosen so that the transition probabilities $w_i(\sigma_i)$ depend on the spin value of $\sigma_i$ as well as on the values of its 
nearest neighbors. 

Using Eqs. (\ref{master}) to (\ref{gamma}) we obtain after some algebraic manipulation the dynamical equations for $m_i(t)$ and 
$r_{ij}(t)$
\begin{eqnarray}
\frac{dm_i(t)}{dt} &=& - \frac{m_i(t)}{\tau} + \frac{1}{\tau} \sum_{\sigma} \tanh{(\Delta_{i})} P_i(\sigma,t)  \ ,
\label{mean}
\\
\frac{dr_{ij}(t)}{dt} &=& - \frac{2 r_{ij}(t)}{\tau} 
\label{pair} \\ 
&+& \frac{1}{\tau} \sum_{\sigma} \left[  \sigma_i \tanh{(\Delta_{j})} + \sigma_j \tanh{(\Delta_{i})} \right] P_{ij}(\sigma,t) \ .
\nonumber
\end{eqnarray}
In principle, our task is to integrate these equations to obtain the time evolution of the variables of interest. 
Within the Bethe-Peierls approximation \cite{Huang}, we assume that only the correlations between pairs of nearest neighbors are nonzero 
and write the following equations for $P_i(\sigma)$
\begin{eqnarray}
P_i(\sigma) =   \frac{\prod_{\alpha}^{N_i} P(\sigma_i,\sigma_{\alpha})}{P^{N_i - 1} (\sigma_i) } \ ,
\label{prob1}
\end{eqnarray}
where $N_i$ is the coordination number of the lattice site $i$ and for $P_{ij}(\sigma)$
\begin{eqnarray}
P_{ij}(\sigma) &=&   \frac{  P(\sigma_i,\sigma_j) }{  P^{N_i - 1} (\sigma_i) P^{N_j - 1} (\sigma_j) } 
\label{prob2} \\
&\times&  \prod_{\alpha}^{N_i} P(\sigma_i,\sigma_{\alpha})    \prod_{\beta}^{N_j} P(\sigma_i,\sigma_{\beta}) \ .
\nonumber
\end{eqnarray}
It is important to notice that $N_i$ is different for the surface spins.
$P(\sigma_i)$ and $P(\sigma_i,\sigma_j)$ are obtained from expression (\ref{prob}) taking as nonzero only the nearest neighbor correlation functions
\begin{eqnarray}
P(\sigma_i) = \frac{1}{2} \left( 1 + m_i \sigma_i \right) \ \ \ \ \ \ \ \ \ \ \  \ , \\
P(\sigma_i,\sigma_j) = \frac{1}{4} \left( 1+ m_i\sigma_i + m_j\sigma_j + r_{ij}\sigma_i\sigma_j \right) \ .
\label{p1p2}
\end{eqnarray}

Substituting  expressions (\ref{prob1}) and (\ref{prob2})  into the dynamical Eqs. (\ref{mean}) and (\ref{pair}) yields, after a lengthy but straightforward 
manipulation, a closed system of equations for the spin magnetizations ($m_i(t)$) and nearest neighbor pair correlation functions ($r_{ij}(t)$) that can 
be solved numerically
\begin{eqnarray}
& & \tau \frac{dm_i(t)}{dt} + m_i(t) =  \nonumber \\
& & \sum_{k=0}^{N_i} \left\{  \frac{t_{6-k}}{x^5_i} \widehat{\mathcal{V}}^k Z_i(\{N_i\}) + \frac{t_k}{y^5_i}
\widehat{\mathcal{U}}^k W_i(\{N_i\}) \right\} \ ,
\label{eq1}
\end{eqnarray}
\begin{eqnarray}
\tau \frac{dr_{ij}(t)}{dt}  + 2 r_{ij}(t) = [\xi_i]_j + [\xi_j]_i \ ,
\label{eq2}
\end{eqnarray}
with 
\begin{eqnarray}
[\xi_i]_j &=& \sum_{k=0}^{N_j-1} \left\{ [z_i]_j t_{6-k} - [v_i]_j t_{5-k} \right\} \frac{ \widehat{\mathcal{V}}^k Z_j(\{N_j\}) }{ x^5_j} 
\nonumber \\ 
&+& \sum_{k=0}^{N_j-1} \left\{ [u_i]_j t_{1+k} - [w_i]_j t_{k} \right\} \frac{ \widehat{\mathcal{U}}^k  W_j(\{N_j\}) }{ y^5_j} \ ,
\label{xi}
\end{eqnarray}
and $[\xi_j]_i$ obtained by the simple exchange of the indices $i$ and $j$ in Equation (\ref{xi}).  Refer to the appendix for a definition of the symbols.
We also define the products
\begin{eqnarray}
Z_j(\{N_j\}) &=& \prod_{\alpha}^{N_j} \ [z_{\alpha}]_j  \\
W_j(\{N_j\}) &=& \prod_{\alpha}^{N_j} \ [w_{\alpha}]_j \ ,
\end{eqnarray}
upon which act the operators
\begin{eqnarray}
\widehat{\mathcal{V}}^k &\equiv&  \sum_{\alpha_1< \ldots < \alpha_k}  
\frac{[v_{\alpha_1}]_j \ldots [v_{\alpha_k}]_j}{[z_{\alpha_1}]_j \ldots [z_{\alpha_k}]_j }  \\
\widehat{\mathcal{U}}^k &\equiv&  \sum_{\alpha_1< \ldots < \alpha_k}  
\frac{[u_{\alpha_1}]_j \ldots [u_{\alpha_k}]_j}{[w_{\alpha_1}]_j \ldots [w_{\alpha_k}]_j}  \ ,
\label{operator} 
\end{eqnarray}
that generate all the possible configurations of the cluster. 
In the case of Eq. (\ref{eq1}) the independent indices $\alpha_k$  stand for all the nearest neighbors ($1 \le \alpha_k \le 6$) of a given spin $\sigma_j$, 
as depicted in Fig. 2(a). 
When calculating the pair correlation $r_{ij} = \langle \sigma_i \sigma_j \rangle$, to avoid counting the pair twice $1 \le \alpha_k \le 5$, as 
shown in Fig. 2(b) for spins $\sigma_i$ and $\sigma_j$. For $k=0$, $\widehat{\mathcal{V}}^0 = 1$ and $\widehat{\mathcal{U}}^0=1$.
The parameter $t_k$ is related to the transition rate of a spin
\begin{eqnarray}
t_k = \tanh{ \left( \frac{2J(k-3) + H}{k_B T} \right) } \ .
\end{eqnarray}
The results comprehending Eqs. (\ref{eq1}) throughout (\ref{operator}) correspond to the case $J_{ij} = J$. To account for the more complex case of 
different $J_{ij}$ in heterogeneous nanoparticles $\widehat{\mathcal{V}}^k$ and $\widehat{\mathcal{U}}^k$ have to be modified to incorporate a 
generalized  parameter $t_k$. This is done in the appendix.

\section{Results and Discussions}

In this section we describe the properties of the model and show that it accounts for many of the phenomena presented
by magnetic nanoparticle systems. 

We start by discussing the properties of the hysteresis of homogeneous ferromagnetic nanoparticles, as yielded by the present model. For this 
purpose consider Figure 3(a) where it is shown hysteresis loops for 3 different nanoparticle sizes $R$ = 1 , 2 and 11. $R$ is the 
number of shells and can be associated with the radius of the nanoparticle. We define the total magnetization of the nanoparticle 
$M \equiv \sum_i m_i(t)/N$, where the sum runs over all the spins of the nanoparticle ($N$).
The smallest nanoparticle ($R$ = 1) has vanishing coercive field and its magnetization is well fitted by a Langevin function, whereas 
the bigger particles ($R$ = 2 through 11) exhibit typical hysteresis loops of ferromagnetic nanoparticles. 
Figure 3(b) shows the dependency of the coercive field ($H_C$) on the number of shells of the nanoparticle, by plotting $H_C$ as a function of 
$R^{-1.6}$, for the reduced temperature $k_B T = 3J$. 
This graph demonstrates that the functional relation $H_C \sim R^{-1.6}$ fits very well the results of our calculations for nanoparticles of size 
$2 \le R \le 10$, with the exception of the $R=1$ and $R=11$ cases. The former is in the superparamagnetic regime \cite{Cullit}, as indicated by Fig. 3(a), whereas
the last already starts to behave as an infinite system. 
The same behavior exhibited by $H_C$ in Fig 3(b) is obtained for the reduced temperature $k_B T = 1.5J$.  
According to our calculations, for $R>10$ the Curie temperature ($T_C$) of the nanoparticle coincides
with the $T_C$ of an infinite cubic system, as yielded by the Bethe-Peierls relation $\coth{[J/k_B T_C]} = 5$ \cite{Plischke}.  
Calculations for a square lattice showed that $H_C \sim L^{-1.5}$ for small clusters, where $L$ stands for its lateral dimension.
Finally, we point out that the mechanism of magnetization reversal that takes place in our nanoparticle is not entirely coherent. 
Because of their 
smaller coordination number, the average magnetization ($m_i$) of the outside spin shells decrease by the action of the reverse field $H$, 
consequently also decreasing the 
magnetization of their neighboring shells and creating a nonuniform magnetization profile. Eventually, when $H$ reaches  $H_C$ all the spins flip together.
This process should be responsible for easier magnetization rotation and lower switching fields than those given by the  Stoner-Wohlfarth model.

In the remainder of this section we consider the properties of heterogeneous nanoparticles, which are more complex systems
consisting of ferromagnetic and antiferromagnetic (or ferrimagnetic) regions coupled with each other. Examples of such systems are oxide-coated cobalt (Co)
or iron (Fe) nanoparticles \cite{Beam}, among many others that exhibit the exchange anisotropy effect \cite{review,book}. 
When cooled in the presence of an external magnetic field the soft ferromagnetic core of the particle aligns with the applied field but its outer 
antiferromagnetic (AF) surface gets ordered only when the 
temperature of the system is lower than $T_N$ (N\'eel temperature) \cite{Cullit}. 
The coupling of the ferromagnetic core with the antiferromagnetic
surface therefore produces a unidirectional exchange anisotropy that shifts the hysteresis loop to higher or lower magnetic fields,
depending on their mutual orientation. To obtain this effect we consider an 8-shell nanoparticle with the following 
characteristics: the exchange interactions in the core ($R$ = 0 to 7) are given by the coupling parameter $J_C = 1$, the outmost shell 
($R=8$) is also ferromagnetic with an exchange constant $J_S = 5$ and the coupling between the $R=7$ and $R=8$ shells is due to an 
antiferromagnetic exchange constant $J_{C-S} = -4$. 
The hysteresis we obtain with our formalism is shown in Figure 4(a), which presents a clear exchange shift. Figure 4(b) shows the magnetization
curves of individual spins within the nanoparticle: in the center of the nanoparticle ($S_0$), in the sixth shell ($S_6$), in the seventh shell 
($S_7$) and in the surface shell ($S_8$). As evidenced by this graph, the surface shell and the layer just below it are 
locked  in a very stable AF configuration, whereas the average magnetization of the inner ferromagnetic shells vary gradually towards the center 
of the particle.
For the sake of comparison, we have also calculated the hysteresis for the same system, but using a simple mean field approach,
whose results are shown in Figures 4(c) and 4(d). For the standard mean field equations each spin experiences only an 
effective field that can be seen as  
an average interaction with all the other spins and, consequently, the exchange shift disappears. Looking at 
the individual spin magnetizations (Fig. 4(d)) it is evident that there is no correlation between the surface and core domains.

Another important experimental technique used to investigate the properties of magnetic nanoparticles is the 
field cooling (FC) and zero field cooling (ZFC) magnetization measurements. For an ensemble of non-interacting homogeneous particles the 
blocking temperature ($T_B$)
is related to the magnetic anisotropy constant and the volume of the nanoparticle \cite{Cullit}.
For a system of interacting nanoparticles see \cite{Prozorov}.
The blocking temperature, where the maximum of the ZFC curve occurs, indicates the point 
in which the thermal energy is comparable to  the average anisotropy barrier in the nanoparticles, and beyond it the magnetization decreases with temperature.  
In the following we show a deblocking behavior that is caused only by the exchange interaction between antiferromagnetic and ferromagnetic
regions in a single nanoparticle. For that purpose consider a 6-shell nanoparticle whose core is predominantly AF, $J_C = -0.5 J_0$ for 
$R$ = 0 to 5, with  a ferromagnetic surface, $J_S = 2 J_0$ for $R$ = 6, which is ferromagnetically coupled with the core, $J_{CS} = 1 J_0$ 
between $R$ = 5 and $R$=6. The energy parameter is $J_0=1$ and the time taken to sweep over the whole temperature interval is $t_s = 10^4 \tau$,
with $\tau$ appearing in Eqs (\ref{mean}) and (\ref{pair}).
The calculations were performed in accordance with the standard FC-ZFC experimental practice \cite{FC-ZFC}.
The obtained FC (solid line) and ZFC (dashed line) magnetization curves are shown in figures 5(a) through 5(f) as a function of the 
reduced temperature, for values of the  applied field that vary from $H=0.05|J_0|$ to $H=0.0001|J_0|$.  
$M$ is the normalized magnetization of the entire nanoparticle, as previously defined.
Because it corresponds to a single particle, $T_B$ is sharply marked and does not exhibit the characteristic dispersion caused by the 
size distribution of the particles.
According to our calculations, $T_B$ is not strongly dependent on $J_C$ however it does depend on $J_{SC}$ and $J_C$. 
Figure 5 shows that  the blocking temperature occurs just about the range $J_{CS} \le k_B T_B \le J_S$, which is confirmed for different 
nanoparticle parameters (size, exchange constant and $t_s$). Despite being observed to take place for nanoparticles of different sizes, the blocking 
temperature effect is very sensitive to the relative strengths between the ferromagnetic and antiferromagnetic exchange constants.
For clarity, Figure 6 displays the relation between the average magnitude of the energies associated with the parallel spin alignment ($U_F + U_H$) 
and anti-parallel spin alignment ($U_{AF}$) configurations within the particle, 
\begin{eqnarray}
U_F  &=&   \langle \sum_{<i,j>} J_S \ \sigma_i \sigma_j + \sum_{<k,l>} J_{CS} \ \sigma_k \sigma_l \rangle \nonumber \\ 
     &=& J_S \sum_{<i,j>} r_{ij} + J_{CS} \sum_{<k,l>} r_{kl} \ , \\
U_H  &=&   H \langle \sum_i \sigma_i \rangle = H \sum_i m_i  \ , \\
U_{AF} &=& \langle \sum_{<i,j>} J_C \ \sigma_i \sigma_j \rangle = J_C \sum_{<i,j>} r_{ij} \ ,
\label{energies}
\end{eqnarray}
in terms of which we define the differential exchange energy $\Delta_{XC} \equiv (U_F + U_H - U_{AF}) / (U_F + U_H + U_{AF})$.
Solid and dashed lines represent FC and ZFC processes, respectively, and the arrows indicate their dynamics.
The contribution of the $U_H$ term to the total energy of the nanoparticle is very small, however it is responsible for the deblocking behavior at $T_B$.
At the high temperature regime the average exchange energy is nonzero, because of a remaining finite short range correlation 
between nearest neighbor pairs. Most of the energy is due to the ferromagnetic ordering.
During the ZFC (dashed line) the AF core overwhelms the ferromagnetic surface and forces the magnetization of the whole nanoparticle to vanish. 
Then, in the subsequent heating, even a very small applied field is enough to trigger the alignment of the ferromagnetic surface, 
which is evidenced by the relative increase of the ferromagnetic energy. 
On the other hand, during the FC process the applied field is able to magnetize the nanoparticle at higher temperatures, before 
the AF correlations of the core become too strong. Thus, the interplay between AF and F exchange energies gives rise to the 
hysteresis in temperature that we observe in the FC-ZFC calculations.

\section{Conclusions}

We have developed a new approach to calculate the classical magnetic properties of single-domain nanoparticles. Treating the nearest neighbor
pair correlation functions explicitly we were able to obtain a simple set of equations for the magnetization at any point in the particle.
In addition, Eqs. (\ref{eq1}) and (\ref{eq2}) can also account for the inclusion of interactions between next nearest neighbors in a straightforward manner, 
by incorporating the corresponding $J_{ij}$ into the operators (\ref{op1}) and (\ref{op2}) and increasing the coordination number $N_j$ of each spin. 
The method can be applied to different lattice geometries and extended to 3D systems. 
The formalism treats correctly the short-range correlations that arise in heterogeneous nanoparticles, which have regions of different 
magnetic properties, e.g., oxidized ferromagnetic particles. Such spin correlations were evidenced by the observation of the exchange shift in hysteresis loops
and by the blocking-deblocking effect in the FC-ZFC calculations. 

\section{Acknowledgment}

The authors thank M.L. Sartorelli for fruitful discussions. Financial support for this work was provided by CNPq-Brazil. 

\section{Appendix}

Here we define some of the notation used in section \ref{theory}. Based on Eq. (\ref{prob}), equations (\ref{eq1}) and (\ref{eq2}) 
are written out in terms of the following simple functions
\begin{eqnarray}
x_i = \frac{1+m_i}{2} \ \ \ \ \ \ \ \ \ &,& \ \ \ \ \ \ \ \ y_i = \frac{1-m_i}{2} \ ;
\label{a1} 
\end{eqnarray}
\begin{eqnarray}
[z_i]_j &=& \frac{1 + m_i + m_j + r_{ij}}{4} \ , 
\end{eqnarray}
\begin{eqnarray} 
[w_i]_j &=& \frac{1 - m_i - m_j + r_{ij}}{4} \ , 
\end{eqnarray}
\begin{eqnarray} 
[v_i]_j &=& \frac{1 - m_i + m_j - r_{ij}}{4} \ , 
\end{eqnarray}
\begin{eqnarray} 
[u_i]_j &=& \frac{1 + m_i - m_j - r_{ij}}{4} \ ,
\label{a3}
\end{eqnarray}
where $[z_i]_j = [z_j]_i$ and $[w_i]_j = [w_j]_i$. 
As already mentioned in the main text, the parameter $t_k$ in Eqs. (\ref{eq1}) and (\ref{eq2}) is related to the transition rate of a spin
\begin{eqnarray}
t_k = \tanh{ \left( \frac{2J(k-3) + H}{k_B T} \right) } \ ,
\end{eqnarray}
for $J_{ij} = J$ and in accordance with the definitions given in the main text. For the case of general exchange constants ($J_{ij}$), the term $t_k$ must be
generalized and incorporated into the operators $\widehat{\mathcal{V}}^k$ and $\widehat{\mathcal{U}}^k$
\begin{eqnarray}
\widehat{\mathcal{V}}^k &\equiv&  \sum_{\alpha_1< \ldots < \alpha_k}  
\frac{[v_{\alpha_1}]_j \ldots [v_{\alpha_k}]_j}{[z_{\alpha_1}]_j \ldots [z_{\alpha_k}]_j }  \times 
\label{op1} \\
& & \tanh{ \left( \frac{ \sum_{\beta=1}^{N_j} J_{\beta j} - 2 ( J_{\alpha_1 j} + \ldots + J_{\alpha_k j} ) + H }{k_B T} \right) } \ ,
\nonumber \\
\nonumber \\
\widehat{\mathcal{U}}^k &\equiv&  \sum_{\alpha_1< \ldots < \alpha_k}  
\frac{[u_{\alpha_1}]_j \ldots [u_{\alpha_k}]_j}{[w_{\alpha_1}]_j \ldots [w_{\alpha_k}]_j} \times
\label{op2} \\ 
& & \tanh{ \left( \frac{ \sum_{\beta=1}^{N_j} J_{\beta j} - 2 ( J_{\alpha_1 j} + \ldots + J_{\alpha_k j} ) + H }{k_B T} \right) }  
\ ,  \nonumber
\end{eqnarray}
so that, for $k=0$ 
\begin{eqnarray}
\widehat{\mathcal{V}}^0 &\equiv& \tanh{ \left( \frac{ \sum_{\beta=1}^{N_j} J_{\beta j}  + H }{k_B T} \right) }       \\
\widehat{\mathcal{U}}^0 &\equiv& \tanh{ \left( \frac{ \sum_{\beta=1}^{N_j} J_{\beta j}  + H }{k_B T} \right) }  \ .
\label{special} 
\end{eqnarray}

\end{multicols}

%%%%%%%%%%%%%%%%%%%%%%%%%%%%%%%%%%%%%%   FIGURE   CAPTIONS   %%%%%%%%%%%%%%%%%%%%%%%%%%%%%%%%%%%%%%%%%%%%%%%%%%%%

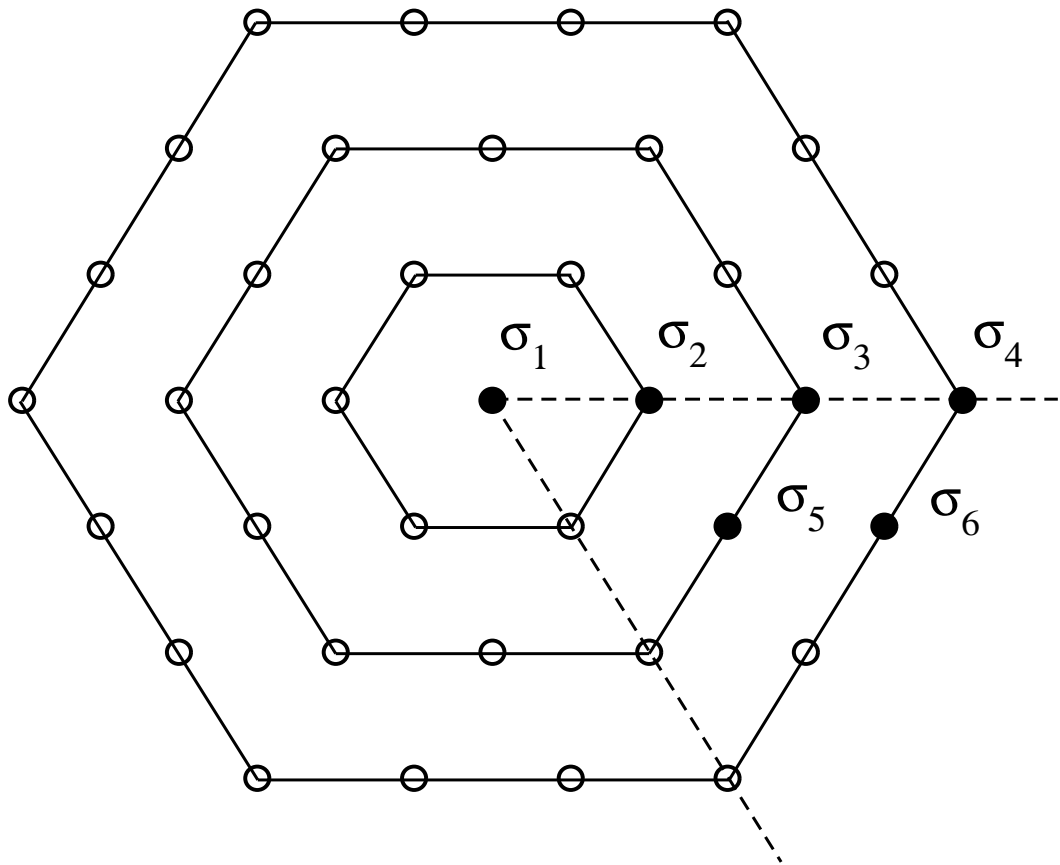
\begin{figure}
\caption{Schematic representation of an hexagonal nanoparticle exhibiting 3 shells of spins (small circles). 
The representative spins that compose the set of equations for this particle are represented by black circles. 
The dashed lines evidence the sixfold symmetry.}
\label{Fig1}
\end{figure}

\begin{figure}
\caption{(a) A cluster formed by spin $\sigma_j$ and its 6 nearest neighbors. (b) A cluster formed by the pair $\sigma_i \sigma_j$ and 
their nearest neighbors.}
\label{1.1}
\end{figure}

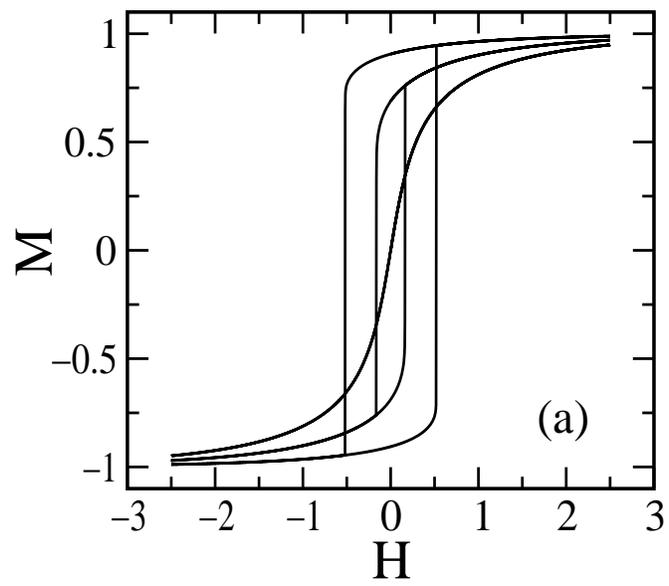
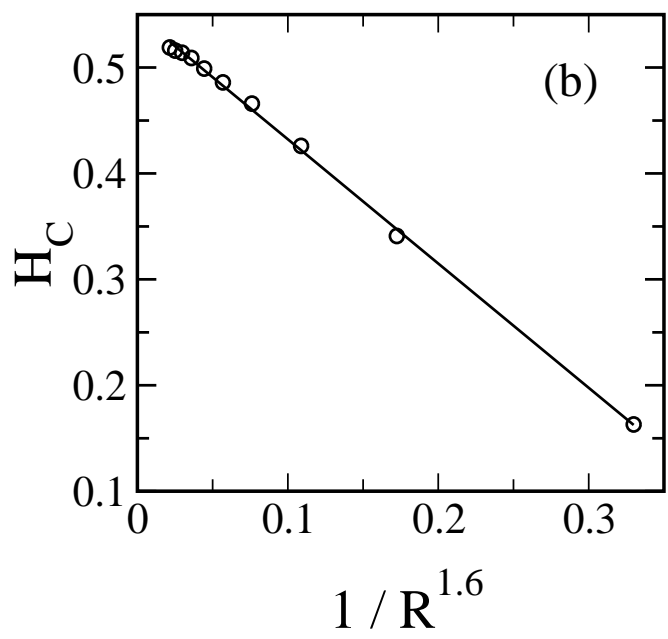
\begin{figure}
\caption{(a) Hysteresis loops for various nanoparticle sizes: $R=1,2$ and 11. The reduced temperature is $T = 3J/k_B$, where $J=1$ is the 
exchange coupling constant and $k_B$ is the Boltzmann constant. 
(b) The coercive field ($H_C$ in units of $J$) plotted as a function of $R^{-1.6}$. From right to left, the dots correspond to 
$R = 2,3,\ldots,11$  and the straight line results from a linear fit to the data.}
\label{Fig2}
\end{figure} 

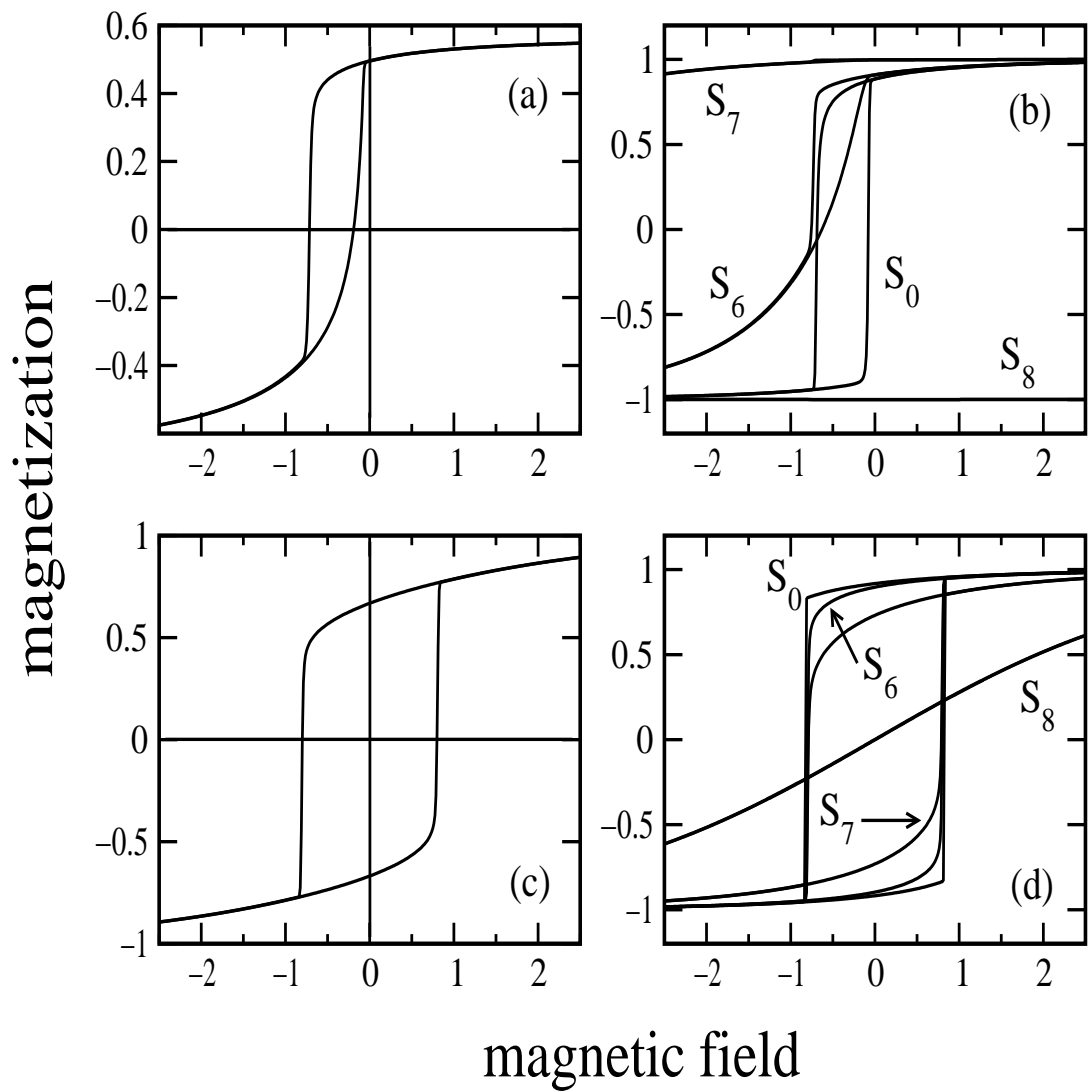
\begin{figure}
\caption{(a) Hysteresis loop for a nanoparticle of radius $R=8$. $J_C = 1$ for $R \le 7$, $J_S = 5$ in the surface $R=8$, and 
$J_{C-S} = -4$ couples the core and surface regions. (b) Individual magnetization curves for the center spin ($S_0$), a
spin in the 6th shell ($S_6$), a spin in the 7th shell ($S_7$) and a spin in the surface shell ($S_8$).
(c) Same parameters as in Figure (a) using a simple mean field approach, the exchange shift disappears. (d) Same as Figure (b) for simple
mean field. The reduced temperature is $k_B T=3.5 J_C$ for all curves.}
\label{Fig3} 
\end{figure}

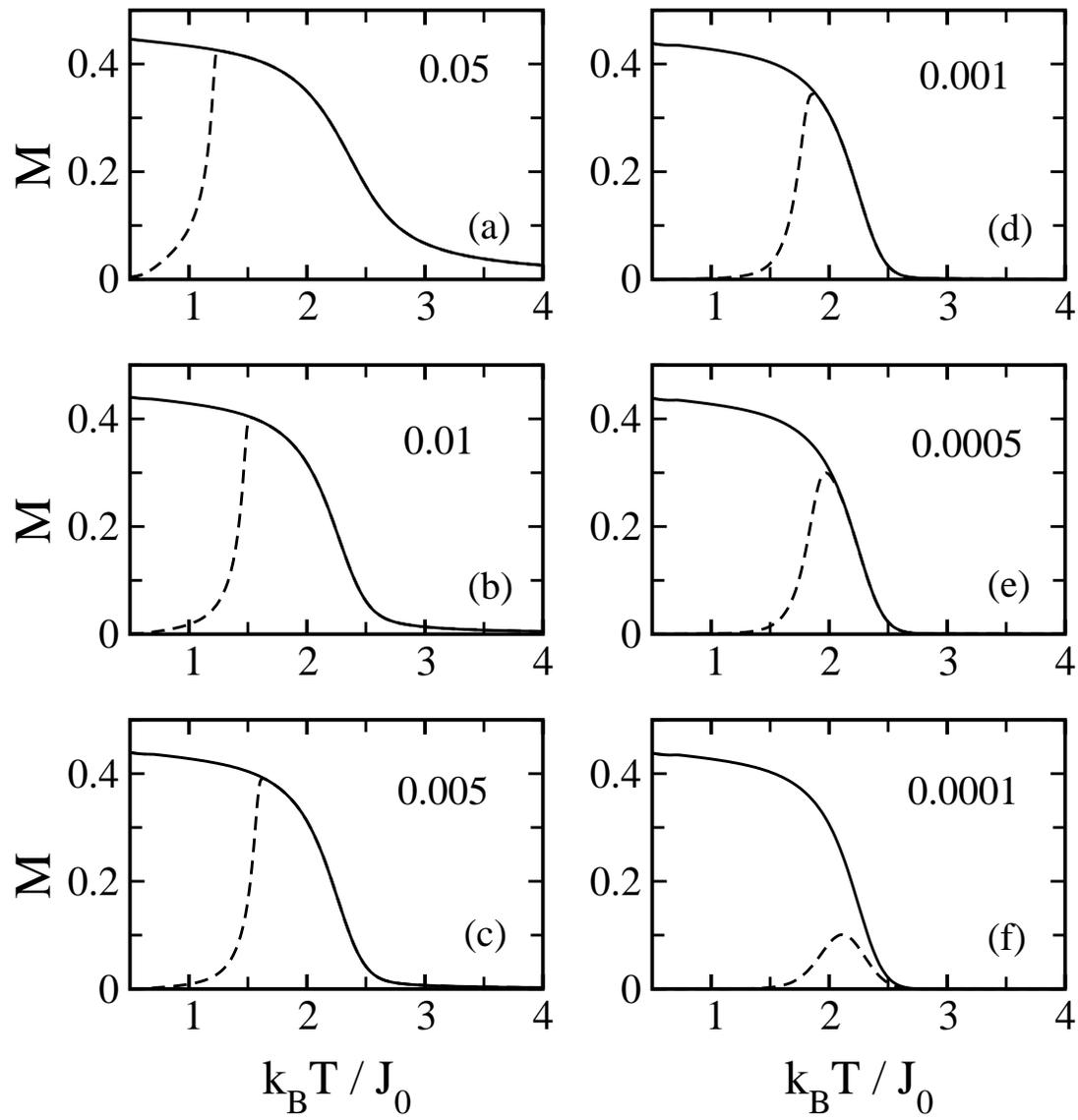
\begin{figure}
\caption{FC (solid) and ZFC (dashed) curves of the total normalized magnetization ($M$) as a function of the reduced temperature 
$k_B T / J_0$. Different values for the applied magnetic field are considered, from $H=0.05|J_0|$ 
in (a) through $H=0.0001|J_0|$ in (f), as indicated in the figures. The nanoparticle consists of $R$ = 6 shells, with an AF core
($J_C = -0.5J_0$, for $0\le R \le 5$), a ferromagnetic surface ($J_S = 2 J_0$ within $R=6$) and a ferromagnetic coupling between
core and surface ($J_{CS} = 1 J_0$, between $R=5$ and $R=6$). }
\label{Fig4}
\end{figure}

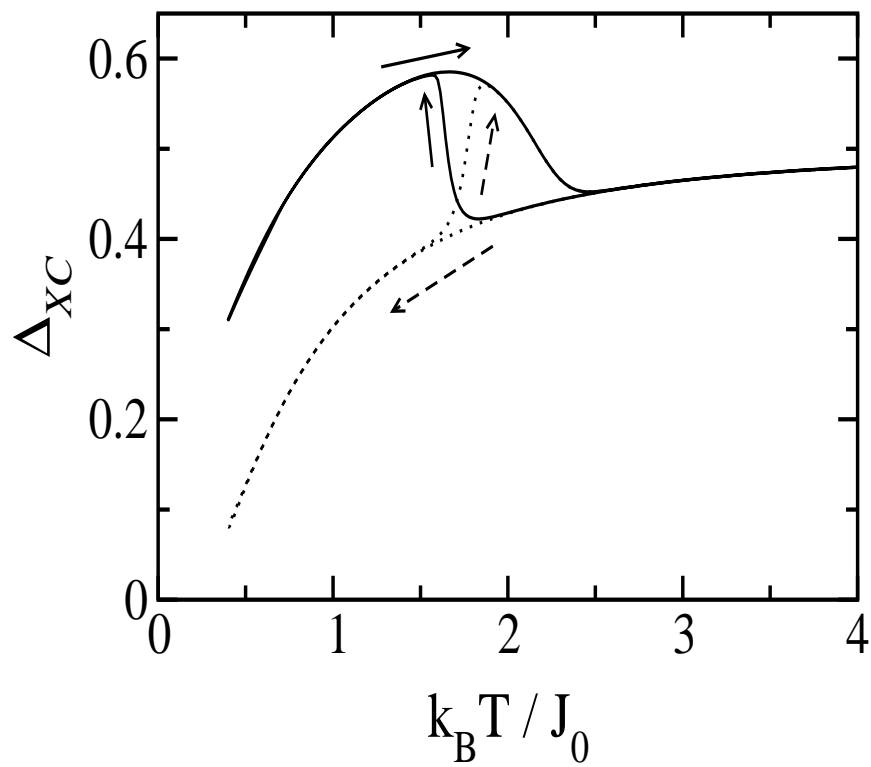
\begin{figure}
\caption{Differential exchange energy $\Delta_{XC} \equiv (U_F + U_H - U_{AF}) / (U_F + U_H + U_{AF})$ as a function of the 
reduced temperature $k_B T / J_0$. The applied field is $H = 0.001 |J_0|$. }
\label{Fig5}
\end{figure}

%%%%%%%%%%%%%%%%%%%%%%%%%%%%%%%%%%%%%%%%%%%%%%%%%%%%%%%%%%%%%%%%%%%%%%%%%%%%%%%%%%%%%%%%%%%%%%%%%%%%%%%%%%%%%%%%%%%%%

\begin{center}
\epsfig{file=fig1.eps,height=15cm,width=14cm}
\end{center}

\begin{center}
\epsfig{file=fig2.eps,height=18cm,width=12cm}
\end{center}

\begin{center}
\epsfig{file=fig3.eps,height=16.5cm,width=12cm}
\end{center}

\begin{center}
\epsfig{file=fig4.eps,height=18cm,width=15cm}
\end{center}

\begin{center}
\epsfig{file=fig5.eps,height=20cm,width=15cm}
\end{center}

\begin{center}
\epsfig{file=fig6.eps,height=13cm,width=14cm}
\end{center}


\begin{references}

\bibitem{Stoner} E.C. Stoner and E.P. Wohlfarth, Philos. Trans. R. Soc. London, Ser. A {\bf 240}, 599 (1948);
reprinted in IEEE Trans. Magn. {\bf 27}, 3475 (1991).

\bibitem{Salah} S.A. Makhlouf, F.T. Parker, F.E. Spada, and A.E. Berkowitz, J. Appl. Phys. {\bf 81}, 5561 (1997);
R.H. Kodama, S.A. Makhlouf, and A.E. Berkowitz, Phys. Rev. Lett. {\bf 79}, 1393 (1997).

\bibitem{Beam} W.H. Meiklejohn and C.P. Bean, Phys. Rev. {\bf 105}, 904 (1957); S. Banerjee, S. Roy, J.W. Chen, and 
C. Chakravorty, J. Mag. Mat. {\bf 219}, 45 (2000).

\bibitem{Kodama} R.H. Kodama, A.E. Berkowitz, E.J. McNiff Jr, and S. Foner, Phys. Rev. Lett. {\bf 77}, 394 (1996); 
B. Martinez, X. Obradors, Ll. Balcells, A. Rouanet, and C. Monty, Phys. Rev. Lett. {\bf 80}, 181 (1997).

\bibitem{Prozorov} R. Prozorov, Y. Yeshurun, T. Prozorov, and A. Gedanken, Phys. Rev. B {\bf 59}, 6956 (1999).

\bibitem{Wernsdorfer1} W. Wernsdorfer, E.B. Orozco, K. Hasselbach, A. Benoit, D. Mailly, O. Kubo, H. Nakano, and B. Barbara,
Phys. Rev. Lett. {\bf 79}, 4014 (1997); E. Bonet, W. Wernsdorfer, B. Barbara, A. Benoit, D. Mailly, and A. Thiaville, Phys. Rev. Lett.
{\bf 83}, 4188 (1999).

\bibitem{Wernsdorfer2} M. Jamet, W. Wernsdorfer, C. Thirion, D. Mailly, V. Dupuis, P. M\'elinon, and A. Per\'ez, Phys. Rev. Lett. 
{\bf 86}, 4676 (2001).

\bibitem{Kirschner} J. Shen, R. Skomski, M. Klaua, H. Jenniches, S.S. Manoharan, and J. Kirschner, Phys. Rev. B {\bf 56}, 2340 (1997).

\bibitem{Plischke} M. Plischke and B. Bergersen, {\it Equilibrium Statistical Physics}, 2nd edition (World Scientific, Singapore,1994).

\bibitem{Glauber} R.J. Glauber, J. Math. Phys. {\bf 4}, 294 (1963).

\bibitem{Huang} K. Huang, Statistical Physics, 2nd edition (Wiley, New York,1987).

\bibitem{Cullit} B.D. Cullit, {\it Introduction to Magnetic Materials} (Addison-Wesley,London,1972).

\bibitem{review} W.H. Meiklejohn, J. Appl. Phys. {\bf 33}, 1328 (1962).

\bibitem{book} R.C. O'Handley, {\it Modern Magnetic Materials: Principles and Applications} (Wiley \& Sons, New York, 1999).

\bibitem{FC-ZFC} R. Sappey, E. Vincent, N. Hadacek, F. Chaput, J.P. Boilot, and D. Zins, Phys. Rev. B {\bf 56}, 14551 (1997).


\end{references}
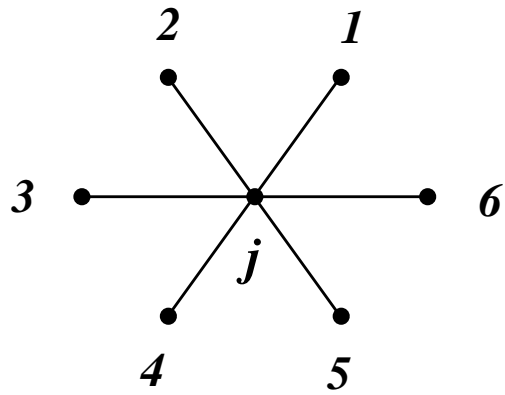
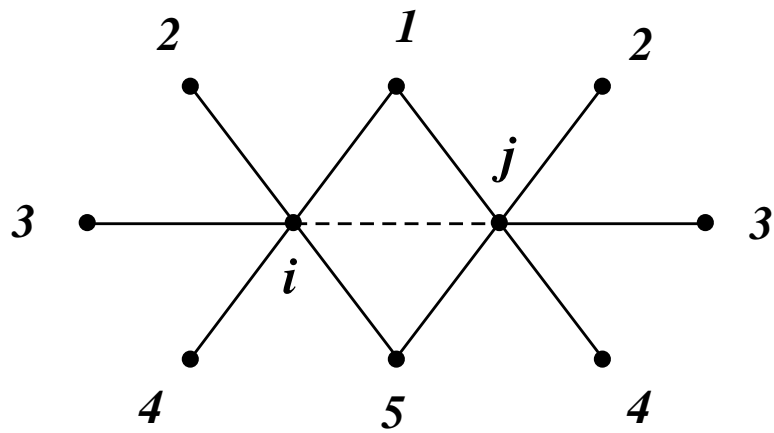
\end{document}